# On the Extraction of Exchange Parameters in Transition-Metal Perovskites

A. Furrer[1], A. Podlesnyak[2], and K. W. Krämer[3]

[1] Laboratory for Neutron Scattering, Paul Scherrer Institut, CH-5232 Villigen PSI, Switzerland

[2] Quantum Condensed Matter Division, Oak Ridge National Laboratory, Oak Ridge, Tennessee 37831-6473, USA

[3] Department of Chemistry and Biochemistry, University of Bern, CH-3012 Bern, Switzerland

**Abstract:**

The extraction of exchange parameters from measured spin-wave dispersion relations has severe limitations particularly for magnetic compounds such as the transition-metal perovskites, where the nearest-neighbor exchange parameter usually dominates the couplings between the further-distant-neighbor spins. Very precise exchange parameters beyond the nearest-neighbor spins can be obtained by neutron spectroscopic investigations of the magnetic excitation spectra of isolated multimers in magnetically diluted compounds. This is exemplified for manganese trimers in the mixed three- and two-dimensional perovskite compounds $KMn_xZn_{1-x}F_3$ and $K_2Mn_xZn_{1-x}F_4$, respectively. It is shown that the small exchange couplings between the second-nearest and the third-nearest neighboring spins can be determined unambiguously and with equal precision as the dominating nearest-neighbor exchange coupling.

# I. INTRODUCTION

Many of the characteristic physical properties of magnetic compounds are determined by their magnetic excitation spectra which reflect the nature of the fundamental magnetic interactions between the spins. A widely used approach to describe the spin interactions is the Heisenberg-Dirac-Van Vleck (HDVV) Hamiltonian [1-3]

$$H = \sum_{i,j} J_{ij} \mathbf{s_i} \cdot \mathbf{s_j} \, , \qquad (1)$$

where $\mathbf{s_i}$ is a spin operator and $J_{ij}$ the exchange parameter coupling the magnetic ions at sites i and j. In the literature the HDVV Hamiltonian is often described with prefactors of -1 or -2, in contrast to the convention adopted here.

Exchange interactions are usually extracted from the spin-wave dispersion which can be measured by inelastic neutron scattering (INS) throughout the Brillouin zone. However, this experimental method has certain limitations. The exchange parameters between nearest-neighbor spins and further-distant-neighbor spins often have similar effects on the dispersion relation and intensity dependence, thus they cannot be determined independently from the data without additional constraints as convincingly demonstrated, *e.g.*, for the spin-wave dispersion of the high-$T_c$ parent compound $La_2CuO_4$ [4]. Moreover, for some manganese compounds associated with colossal magnetoresitance and/or multiferroic behavior, the further-distant-neighbor exchange parameters often suffer from rather large error bars [5] or pose yet unexplained features such as unusually large values [6-8] or could not be extracted at all [9]. Here we will show that these drawbacks can be overcome by INS studies of the spin-excitation spectra of isolated multimers of magnetic ions in magnetically diluted compounds.

The experimental and theoretical aspects associated with isolated n-mers of magnetic ions (often denoted as magnetic clusters) were recently reviewed by Furrer and Waldmann [10]. As long as the number n of spins is reasonably small, exact analytical solutions of the spin Hamiltonian can be obtained. Accordingly, small clusters up to four magnetic ions turn out to be ideal model systems to examine the fundamental magnetic interactions. This type of research started about six decades ago, and numerous investigations showed that the usually dominating HDVV exchange is often complemented by anisotropic and/or higher-order interactions. However, despite this long history, the research field is not yet exhausted, but continuously increases due to the ongoing improvements of the experimental equipment. More specifically, modern neutron spectrometers have provided unprecedented energy resolutions which allow to detect hitherto hidden splittings of lines associated with n-mer excitations. Typical examples include splittings of the lowest-lying singlet-triplet dimer transitions into several components observed, e.g., for $Cu^{2+}$ dimers in $SrCu_2(BO_3)_2$ due to the Dzyaloshinski-Moriya interaction [11] as well as for some $Mn^{2+}$ dimers due to the single-ion anisotropy [12,13] and due to structural inhomogeneities [12,14,15]. While most studies along these lines so far dealt with dimer systems, we present here a new example associated with trimer transitions.

The present work reports on a high-resolution INS study of the spin-excitation spectra of isolated multimers of $Mn^{2+}$ ions randomly substituted for 10% of the nonmagnetic $Zn^{2+}$ ions in the three- and two-dimensional perovskite lattices $KZnF_3$ and $K_2ZnF_4$, respectively. We start with a description of the experimental procedure in Sec. II, followed in Sec. III by a summary of the spin Hamiltonians, energy levels, neutron cross-sections, and cluster probabilities for $Mn^{2+}$ dimers and trimers. Sec. IV presents the experimental results and their analyses, with particular emphasis on the trimer excitations which allow a direct extraction of the second- and third-nearest-neighbor exchange interactions. Finally, some conclusions are given in Sec. V.

## II. EXPERIMENT

The synthesis and the characterization of the samples was described in Ref. [12]. $KMn_{0.10}Zn_{0.90}F_3$ crystallizes in the cubic space group *Pm3m* with a lattice parameter a=4.05733(2) Å at T=10 K. $K_2Mn_{0.10}Zn_{0.90}F_4$ crystallizes in the tetragonal space group *I4/mmm* with lattice parameters a=4.04850(5) Å and c=13.07255(23) Å at T=5 K.

The INS experiments were performed with use of the time-of-flight spectrometer CNCS [16] at the spallation neutron source SNS at Oak Ridge National Laboratory. The samples were enclosed in Al cylinders (12 mm diameter, 45 mm height) and placed into a He cryostat to achieve a minimum temperature of T=2 K. The energy of the incoming neutrons was 3 meV, giving rise to energy resolutions of Gaussian shape with full width at half maximum (FWHM)=44 μeV at E≈0.8 meV and FWHM=27 μeV at E≈1.9 meV, where E is the energy transfer in the neutron energy-loss configuration. Data were collected for scattering vectors **Q** with moduli $0.9 \leq Q \leq 1.7$ Å$^{-1}$.

## III. THEORETICAL BACKGROUND

**A. Spin Hamiltonian for a two-dimensional antiferromagnet**

Let us consider the spin Hamiltonian for a two-dimensional antiferromagnet as used, *e.g.*, for the spin-wave analysis of magnetic excitations in $La_2CuO_4$ [4]:

$$H = J_1 \sum_{i,j} \mathbf{s_i} \cdot \mathbf{s_j} + J_2 \sum_{i,k} \mathbf{s_i} \cdot \mathbf{s_k} + J_3 \sum_{i,l} \mathbf{s_i} \cdot \mathbf{s_l}$$
$$+ J_c \sum_{i,j,k,l} \left[ (\mathbf{s_i} \cdot \mathbf{s_j})(\mathbf{s_k} \cdot \mathbf{s_l}) + (\mathbf{s_i} \cdot \mathbf{s_l})(\mathbf{s_k} \cdot \mathbf{s_j}) - (\mathbf{s_i} \cdot \mathbf{s_k})(\mathbf{s_j} \cdot \mathbf{s_l}) \right] ,$$
(2)

where $J_1$, $J_2$, and $J_3$ are the first-, second-, and third-nearest-neighbor exchange parameters, respectively, and $J_c$ is the ring exchange coupling four spins as sketched in the insert of Fig. 1. The corresponding spin-wave energies are given by [4]

$$E(\mathbf{Q}) = 2Z_c(\mathbf{Q})[A^2(\mathbf{Q}) - B^2(\mathbf{Q})]^{1/2} ,$$

$$A(\mathbf{Q}) = J_1 - J_c/2 - (J_2 - J_c/2)(1 - r_h r_k) - J_3[1 - (r_{2h} + r_{2k})/2] ,$$
(3)

$$B(\mathbf{Q}) = (J_1 - J_c/2)(r_h + r_k)/2 , \quad r_x = \cos(2\pi x) .$$

$Z_c(\mathbf{Q})$ is a renormalization factor that describes the effect of quantum fluctuations. For compounds with $s_i \neq 1/2$, Eq. (2) often includes a single-ion anisotropy term which gives rise to an energy gap at the zone center.

Fig. 1 shows the calculated spin-wave dispersion along the xy-direction for different values of the exchange parameters in dimensionless units (using $s_i = 1/2$ and $Z_c(\mathbf{Q}) = 1$). We recognize that all four exchange parameters have similar effects on the dispersion relation. The situation is slightly improved when considering other symmetry directions, but in no case it is possible to arrive at an unambiguous parametrization as emphasized in Ref. [4].

**B. Spin Hamiltonian of magnetic dimers and trimers**

The spin Hamiltonian of dimer excitations reads

$$H = J_1 \mathbf{s}_1 \cdot \mathbf{s}_2 + D\left[\left(s_1^z\right)^2 + \left(s_2^z\right)^2\right]. \tag{4}$$

D is the single-ion anisotropy parameter. The diagonalization of Eq. (4) is based on the dimer states $|S,M\rangle$, where $\mathbf{S}=\mathbf{s}_1+\mathbf{s}_2$ is the total spin and $-S \leq M \leq S$. For antiferromagnetic exchange $J_1$ the ground state is always a singlet (S=0), separated from the first-excited triplet (S=1) by the energy $J_1$. The effect of D is to split the triplet into a doublet and a singlet as shown in Fig. 2(a) for D>0. For the $Mn^{2+}$ compounds studied in the present work this splitting amounts to 6.39D [12,17].

The Hamiltonian for spin excitations associated with equilateral trimers is given by

$$H = J_1\left(\mathbf{s}_1 \cdot \mathbf{s}_2 + \mathbf{s}_2 \cdot \mathbf{s}_3\right) + J'\mathbf{s}_1 \cdot \mathbf{s}_3 + D\left[\left(s_1^z\right)^2 + \left(s_2^z\right)^2 + \left(s_3^z\right)^2\right]. \tag{5}$$

J' is the next-nearest-neighbor exchange parameter which corresponds either to $J_2$ or $J_3$ of Eq. (2) depending on the geometrical trimer configuration (see inserts in Figs. 1 and 3). The trimer states $|S_{13},S,M\rangle$ are defined by the spin coupling scheme $\mathbf{S_{13}}=\mathbf{s_1}+\mathbf{s_3}$, $\mathbf{S}=\mathbf{s_1}+\mathbf{s_2}+\mathbf{s_3}$ with $0 \leq S_{13} \leq 2s_i$ and $|S_{13}-s_i| \leq S \leq (S_{13}+s_i)$, and $-S \leq M \leq S$. For antiferromagnetic exchange and $Mn^{2+}$ ions ($s_i=5/2$) the ground state is the sextet $|5,5/2,M\rangle$, separated from the lowest excited quartet state $|4,3/2,M\rangle$ by the energy $2.5J_1-5J'$. The effect of D is to split these states into doublets as shown in Fig. 2(b) for D>0. For the $Mn^{2+}$ compounds studied in the present work these splittings amount to about 1.5D, *i.e.*, they are considerably smaller than for the dimer case, thus they can usually not be resolved in INS experiments.

**C. Neutron cross-section for dimer and trimer transitions**

Transitions between different dimer and trimer states can be directly measured by INS experiments according to the selection rules $\Delta S=0,\pm 1$ and $\Delta M=0,\pm 1$; for trimer transitions the additional selection rule $\Delta S_{13}=0,\pm 1$ holds. For spin dimers and polycrystalline material the neutron cross-section for a transition from the initial state $|S,M\rangle$ to the final state $|S',M'\rangle$ was worked out in detail in Ref. [18]. For the neutron cross-section of trimer transitions $|S,S_{13},M\rangle \rightarrow |S',S'_{13},M'\rangle$ we refer to Ref. [19]. Since in the present work we are only interested in transitions with $\Delta S_{13}=\pm 1$, the trimer cross-section including the summation $\Sigma_{M,M'}$ reduces to

$$\frac{d^2\sigma}{d\Omega d\omega} = \frac{N}{Z}(\gamma r_0)^2 F^2(Q) \exp\left\{-\frac{E(S_{13},S)}{k_B T}\right\} \frac{4}{3}\left[1-\frac{\sin(QR)}{QR}\right] \\ \times \left|\langle S'_{13},S'\|T\|S_{13},S\rangle\right|^2 \delta\{\hbar\omega + E(S_{13},S) - E(S'_{13},S')\} \quad . \tag{6}$$

N is the total number of trimers in the sample, Z the partition function, $\gamma$ a constant with $\gamma=-1.91$, $r_0$ the classical electron radius, F(Q) the magnetic form factor, $k_B$ the Boltzmann factor, T the temperature, $\hbar\omega$ the energy transfer, $E(S_{13},S)$ and $E(S'_{13},S')$ the energy of the initial and final state, respectively, R the distance between the end spins of the trimer, and $\langle S'\|T\|S\rangle$ the reduced transition matrix element defined in Ref. [19]. The structure factor S(Q)=[1-sin(QR)/QR] directly reflects the trimer geometry as illustrated in Fig. 3 for the cases of straight and angled trimers with bond angles of 180° and 90°, respectively. The pronounced phase shift of the damped oscillatory behavior of S(Q) for the two cases is extremely helpful for an unambiguous identification of the particular trimer type.

### D. Multimer probabilities

Multimers of $Mn^{2+}$ ions in the considered compounds occur simply because of the random distribution of $Mn^{2+}$ and $Zn^{2+}$ ions over the sites of the cubic perovskite lattice. For the three-dimensional compound $KMn_xZn_{1-x}F_3$, the probabilities p that a given $Mn^{2+}$ ion is in a particular multimer, are defined by [20]

$$p_M=(1-x)^6 \, , \; p_D=6x(1-x)^{10} \, , \; p_{Ts}=9x^2(1-x)^{14} \, , \; p_{Ta}=24x^2(1-x)^{13} \, , \qquad (7)$$

where the indices M, D, $T_s$, and $T_a$ refer to monomers, dimers, straight trimers, and angled trimers, respectively. For the two-dimensional compound $K_2Mn_xZn_{1-x}F_4$ the corresponding probabilities are

$$p_M=(1-x)^4 \, , \; p_D=4x(1-x)^6 \, , \; p_{Ts}=6x^2(1-x)^8 \, , \; p_{Ta}=8x^2(1-x)^7 \, . \qquad (8)$$

### IV. RESULTS AND DATA ANALYSIS

#### A. $KMn_{0.10}Zn_{0.90}F_3$

Energy spectra of neutrons scattered from $KMn_{0.10}Zn_{0.90}F_3$ at T=2 K are shown in Fig. 4 corresponding to the lowest-lying $Mn^{2+}$ dimer (a) and trimer (b) transitions (see Fig. 2). Since the instrumental energy resolution was optimized for the trimer transition, the fine structure of the dimer transition observed in earlier INS experiments [13] is smeared out. The solid and broken curves in (a) correspond to a superposition of the fine structure lines (folded with the instrumental energy resolution) for both the $|0,0\rangle \rightarrow |1,\pm 1\rangle$ ($D_{\pm 1}$) and the $|0,0\rangle \rightarrow |1,0\rangle$ ($D_0$) transitions reported in Ref. [13], and the only adjustable parameters were a linear background and an overall intensity scaling factor. The agreement between the experimental and the calculated data nicely confirms the model parameters of Eq. (4), namely $J_1$=0.819(4) meV [21] and D=5.3(2) μeV as derived in Ref. [13]. This is important for the analysis of the trimer transition which depends on these parameters in addition to J', see Eq. (5).

The transition shown in Fig. 4(b) includes the response of both the straight ($T_s$) and the angled ($T_a$) $Mn^{2+}$ trimers, with individual weights according to Eq. (7). For J'=0, Eq. (5) predicts the trimer splitting to be at an energy transfer of 2.05(1) meV. However, the data show a significant shift to lower energy transfers, which underlines the existence of appreciable J' interactions. Moreover, the trimer transition exhibits an asymmetric shape, which indicates that the J' interactions for the $T_s$ and $T_a$ trimers are different. We therefore analyzed the data by two Gaussian lines without any constraints in the least-squares fitting procedure, except for fixing the widths of the two lines at equal values. The results are shown as full and broken curves in Fig. 4(b). The intensity ratio $T_a/T_s$ agrees well with the ratio $p_{Ta}/p_{Ts}$=2.96 predicted by Eq. (7). Our identification of the $T_s$ and $T_a$ transitions is furthermore confirmed by the Q-dependence of the intensities based on Eqs. (6) and (7) as illustrated in Fig. 3. The width of the two Gaussian lines with FWHM=147(8) μeV considerably exceeds the instrumental energy resolution (FWHM=27 μeV) due to local structural inhomogeneities as verified for the dimer transitions in Ref. [13].

From the positions of the trimer transitions $T_s$ and $T_a$ we can now derive the exchange parameter J' on the basis of Eq. (5), with the parameters $J_1$ and D fixed at the values of Ref. [13]. We find

$J_2$=0.029(3) meV , $J_3$= 0.011(3) meV ,

*i.e.*, both interactions are antiferromagnetic. For the identification of J' with $J_2$ and $J_3$ corresponding to angled and straight trimers, respectively, we refer to the insert of Fig. 1. The analysis of the spin-wave dispersion measured for $KMnF_3$ gave $J_2$=0.019(4) meV, but the parameter $J_3$ could not be extracted [22]. The smaller size of $J_2$ is due to the larger lattice parameter of $KMnF_3$ (a=4.182 Å) compared to $KMn_{0.10}Zn_{0.90}F_3$ (a=4.057 Å).

## B. $K_2Mn_{0.10}Zn_{0.90}F_4$

Energy spectra of neutrons scattered from $K_2Mn_{0.10}Zn_{0.90}F_4$ at T=2 K are shown in Fig. 5 corresponding to the lowest-lying $Mn^{2+}$ dimer (a) and trimer (b) transitions (see Fig. 2). The data were analyzed in the same way as described in Sec. IV.A. The nice agreement between the experimental and the calculated data in Fig. 5(a) confirms the model parameters of Eq. (4), namely $J_1$=0.826(4) meV [21] and D=5.2(2) μeV as derived in Ref. [13].

The transition shown in Fig. 5(b) results from both the straight ($T_s$) and the angled ($T_a$) $Mn^{2+}$ trimers, with individual weights according to Eq. (8). For J'=0, Eq. (5) predicts the trimer splitting to be at an energy transfer of 2.07(1) meV, however, the existence of appreciable J' interactions shifts the data to lower energy transfers. Moreover, the asymmetric shape of the trimer transition indicates different J' interactions for the $T_s$ and $T_a$ trimers. The results of the least-squares fitting procedure in terms of two Gaussian lines with equal width FWHM=128(7) μeV are shown as full and broken curves in Fig. 5(b). As mentioned in Sec. IV.A, the line broadening is mainly caused by the finestructure effects discussed for the dimer transitions in Ref. [13]. The intensity ratio $T_a/T_s$ agrees well with the ratio $p_{Ta}/p_{Ts}$=1.48 predicted by Eq. (8). Furthermore, our identification of the $T_s$ and $T_a$ transitions is confirmed by the agreement of the Q-dependence of the intensities based on Eqs. (6) and (8).

From the positions of the trimer transitions $T_s$ and $T_a$ we can now derive the exchange parameter J' on the basis of Eq. (5), with the parameters $J_1$ and D fixed at the values of Ref. [13]. We find

$J_2$=0.013(3) meV , $J_3$= 0.032(3) meV ,

*i.e.*, both interactions are antiferromagnetic. The identification of J' with $J_2$ and $J_3$ is explained in Sec. IV.A. The analysis if the spin-wave dispersion measured for $K_2MnF_4$ [23] failed to provide values for $J_2$ and $J_3$.

## V. CONCLUDING REMARKS

By analyzing the lowest lying $Mn^{2+}$ trimer excitations observed for the perovskite compounds $KMn_{0.10}Zn_{0.90}F_3$ and $K_2Mn_{0.10}Zn_{0.90}F_4$, we were able to extract the second-nearest-neighbor and third-nearest-neighbor exchange interactions with a precision similar to the dominating nearest-neighbor exchange coupling. Our experimental method does not suffer from approximations that usually have to be adopted for spin-wave models. In particular, the exchange parameters can be determined without any scaling factors such as the renormalization factor $Z_c(\mathbf{Q})$ of Eq. (2) relevant for magnetic compounds with low spin ($s_i$=1/2 or $s_i$=1).

The application of our experimental method for some manganese perovskites exhibiting colossal magnetoresistance and/or multiferroic behavior may shed light on both the size and the sign of further-distance-neighbor exchange parameters. The analysis of the measured spin-wave dispersion relations resulted in small values of the exchange parameters $J_2$ and $J_3$ (with rather large error bars [5]), but an unusually large value of $J_4 \approx J_1/2$ was reported [6-8]. These findings could be verified or corrected by INS measurements for $Mn^{3+}$ multimers in magnetically diluted compounds. More specifically, the small exchange parameters $J_2$ and $J_3$ can be derived from $Mn^{3+}$ trimer excitations as outlined in the present work, whereas the dominating exchange parameters $J_1$ and $J_4$ are directly accessible through the corresponding $Mn^{3+}$ dimer excitations. In fact, the simultaneous presence of different types of dimer excitations can unambiguously be disentangled as demonstrated by INS experiments performed for the perovskite compound $LaMn_{0.1}Ga_{0.9}O_3$ [24]. Similarly, very recent spin-wave data obtained for the layered ferroelectric compound $LuFe_2O_4$ resulted in two different, but statistically equivalent sets of five nearest-neighbor exchange interactions within and between the monolayers [25], which could easily be discriminated from each other by studying the corresponding $Fe^{2+}/Fe^{3+}$ dimer excitations.

## ACKNOWLEDGMENTS

The assistance of D. Biner (University of Bern) in the synthesis of the samples is gratefully acknowledged. Research at Oak Ridge National Laboratory's Spallation Neutron Source was supported by the Scientific User Facilities Division, Office of Basic Energy Sciences, US Department of Energy.

**FIGURE CAPTIONS**

FIG. 1 (Color online) Spin-wave dispersion of a two-dimensional antiferromagnet along the xy-direction calculated from Eq. (2) for different values of the exchange parameters in dimensionless units (using $s_i=1/2$ and $Z_c(\mathbf{Q})=1$). The insert shows the exchange parameters $J_n$ coupling the up- and down-spins marked by open and filled circles, respectively.

FIG. 2 (Color online) Low-energy level splittings of $Mn^{2+}$ dimers (a) and trimers (b). The trimer splittings are enhanced by a factor of three. The transitions relevant for the present work are marked by arrows. The five allowed trimer ground-state transitions (T) resulting from the anisotropy splitting can usually not be resolved in INS experiments, in contrast to the two ground-state dimer transitions ($D_{\pm 1}$ and $D_0$).

FIG. 3 (Color online) Q-dependence of the structure factor S(Q) for $Mn^{2+}$ trimer transitions with $\Delta S_{13}=\pm 1$ for straight ($T_s$) and angled ($T_a$) trimer geometries shown in the insert. The data correspond to the results obtained for $KMn_{0.10}Zn_{0.90}F_3$.

FIG. 4 (Color online) Energy spectra of neutrons scattered from $KMn_{0.10}Zn_{0.90}F_3$ at T=2 K. The lines are the results of a least-squares fitting procedure explained in the text. The arrows mark the positions of the $Mn^{2+}$ dimer (a) and trimer (b) transitions.

FIG. 5 (Color online) Energy spectra of neutrons scattered from $K_2Mn_{0.10}Zn_{0.90}F_4$ at T=2 K. The lines are the results of a least-squares fitting procedure explained in the text. The arrows mark the positions of the $Mn^{2+}$ dimer (a) and trimer (b) transitions.

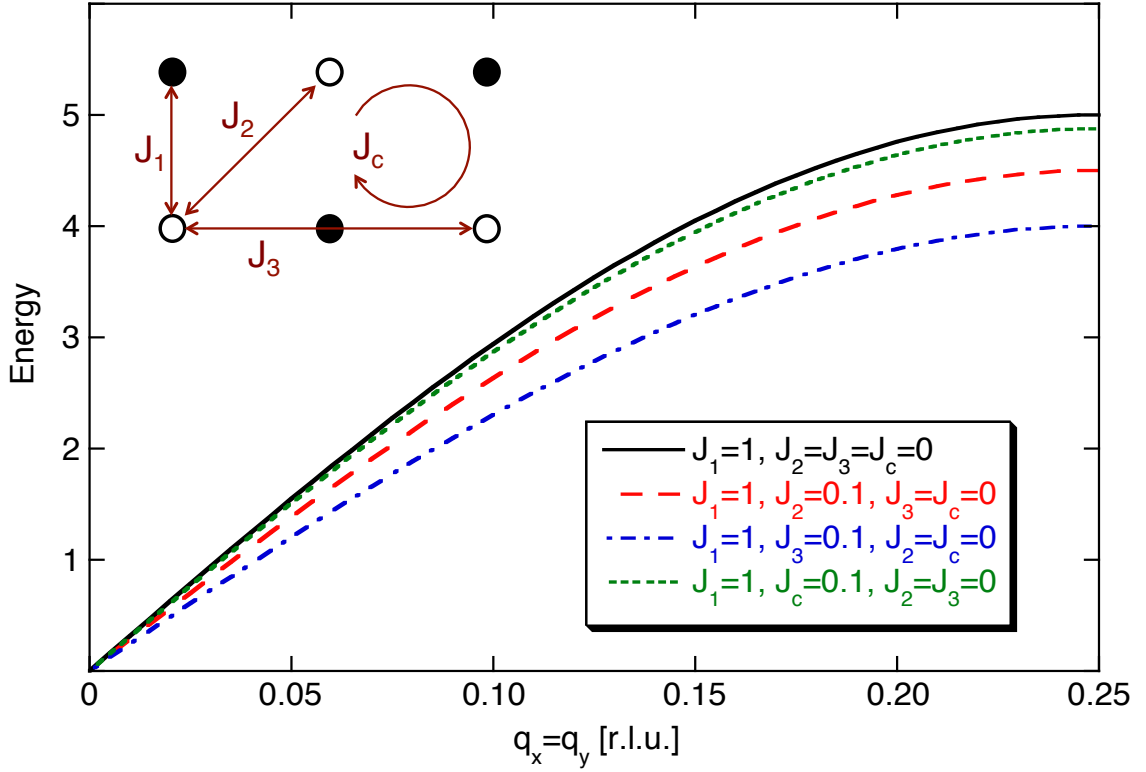

FIG. 1 (Color online) Spin-wave dispersion of a two-dimensional antiferromagnet along the xy-direction calculated from Eq. (2) for different values of the exchange parameters in dimensionless units (using $s_i=1/2$ and $Z_c(\mathbf{Q})=1$). The insert shows the exchange parameters $J_n$ coupling the up- and down-spins marked by open and filled circles, respectively.

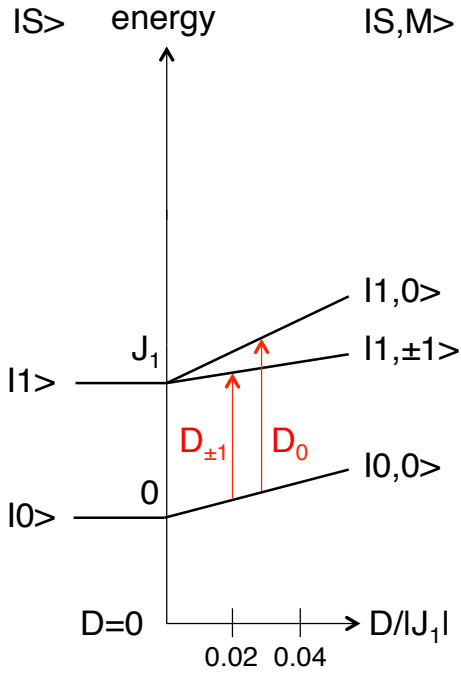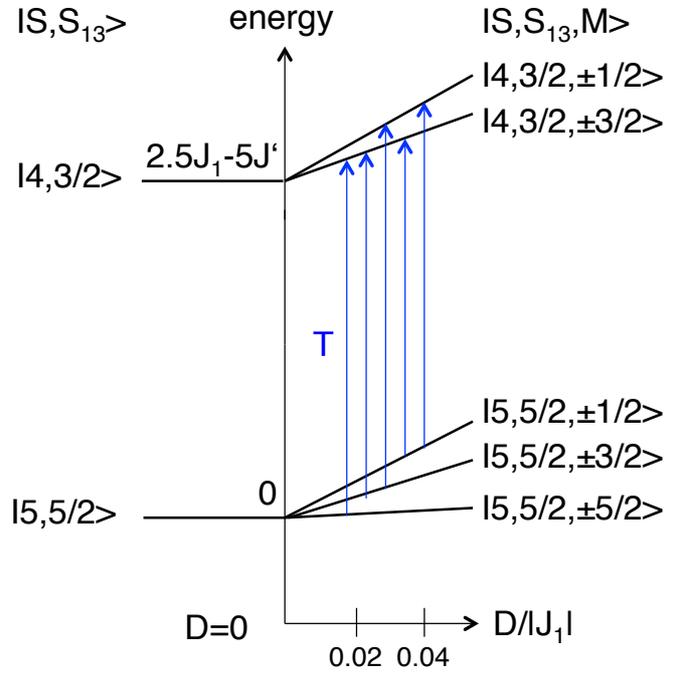

FIG. 2 (Color online) Low-energy level splittings of $Mn^{2+}$ dimers (a) and trimers (b). The trimer splittings are enhanced by a factor of three. The transitions relevant for the present work are marked by arrows. The five allowed trimer ground-state transitions (T) resulting from the anisotropy splitting can usually not be resolved in INS experiments, in contrast to the two ground-state dimer transitions ($D_{\pm 1}$ and $D_0$).

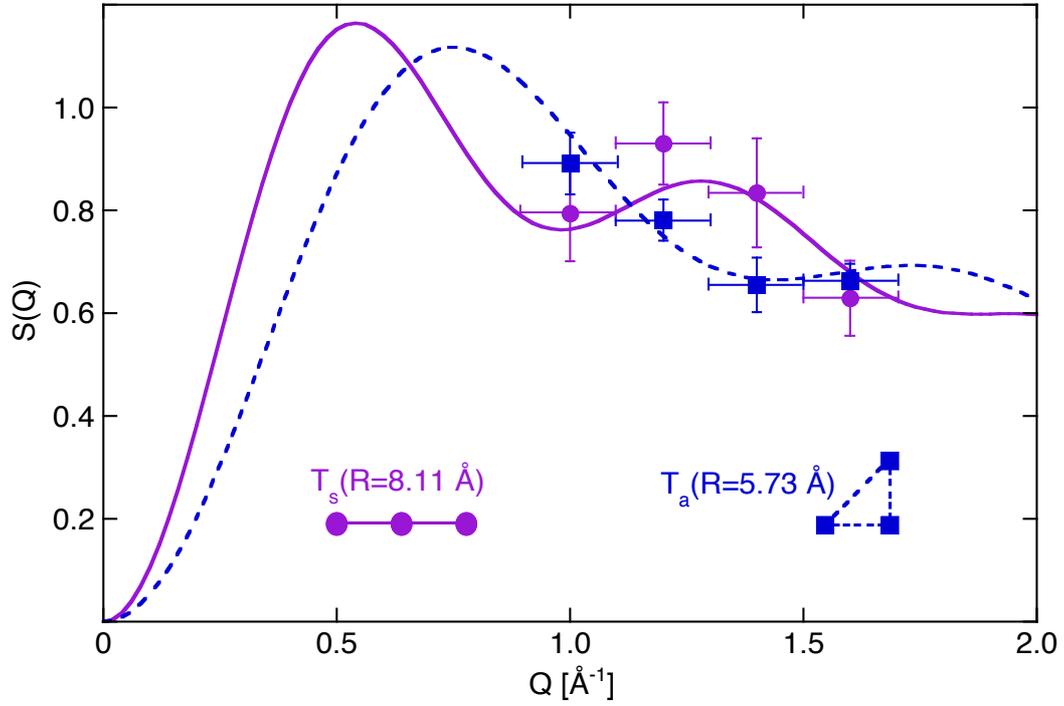

FIG. 3 (Color online) Q-dependence of the structure factor S(Q) for $Mn^{2+}$ trimer transitions with $\Delta S_{13}=\pm 1$ for straight ($T_s$) and angled ($T_a$) trimer geometries shown in the insert. The data correspond to the results obtained for $KMn_{0.10}Zn_{0.90}F_3$.

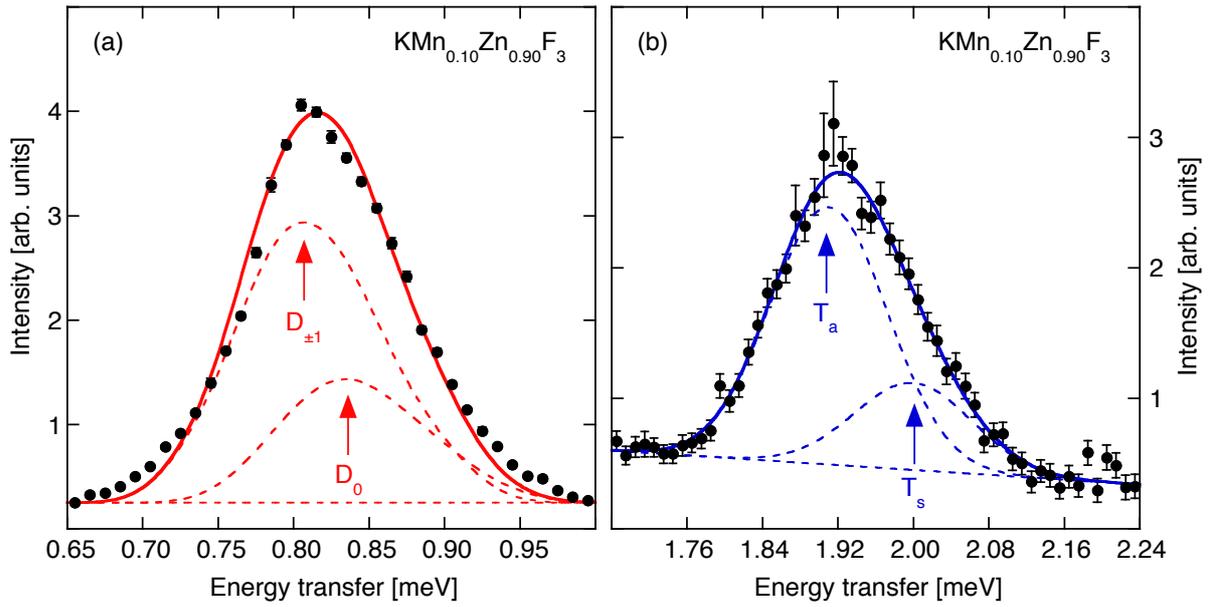

FIG. 4 (Color online) Energy spectra of neutrons scattered from KMn$_{0.10}$Zn$_{0.90}$F$_3$ at T=2 K. The lines are the results of a least-squares fitting procedure explained in the text. The arrows mark the positions of the Mn$^{2+}$ dimer (a) and trimer (b) transitions.

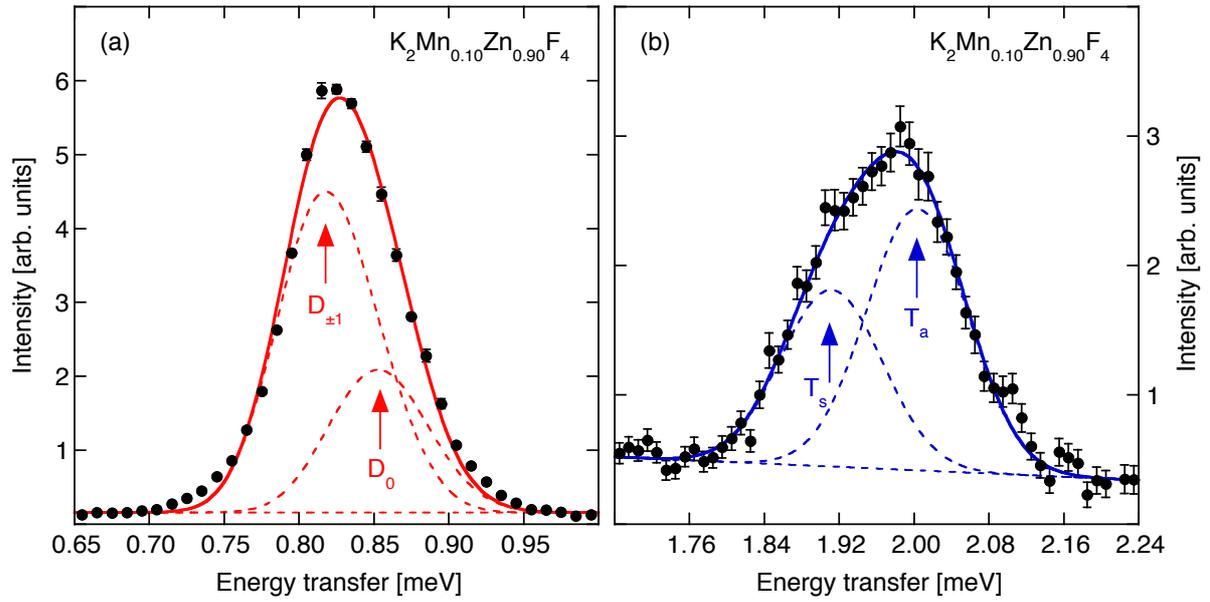

FIG. 5 (Color online) Energy spectra of neutrons scattered from $K_2Mn_{0.10}Zn_{0.90}F_4$ at T=2 K. The lines are the results of a least-squares fitting procedure explained in the text. The arrows mark the positions of the $Mn^{2+}$ dimer (a) and trimer (b) transitions.